\documentclass[12pt]{iopart}

\usepackage{graphicx}
\usepackage{ulem}
\usepackage{textcomp}

\begin{document}

\paper[Grain boundary characteristics of oxypnictide NdFeAs(O,F) superconductors]{Grain boundary characteristics of oxypnictide NdFeAs(O,F) superconductors}

\author{Kazumasa Iida$^{1,2}$, Taito Omura$^{2}$, Takuya Matsumoto$^{1}$, Takafumi Hatano$^{1,2}$, Hiroshi Ikuta$^{1,2}$}

\address{$^1$Department of Materials Physics, Nagoya University, Furo-cho, Chikusa-ku, Nagoya 464-8603, Japan\\$^2$Department of Crystalline Materials Science, Nagoya University, Furo-cho, Chikusa-ku, Nagoya 464-8603, Japan}
\ead{iida@mp.pse.nagoya-u.ac.jp}
\begin{abstract}
We have systematically investigated the grain boundary angle ($\theta_{\rm GB}$) dependence of transport properties for NdFeAs(O,F) fabricated on [001]-tilt symmetric MgO bicrystal substrates. In our previous study, NdFeAs(O,F) bicrystal films showed a weak-link behaviour even at a $\theta_{\rm GB}$ of 6$^\circ$. However, this was caused by an extrinsic effect originating from the damage to both NdFeAs(O,F) and MgO substrate by excess F-diffusion along the grain boundary. To investigate the intrinsic nature of grain boundaries, we minimized the damage to NdFeAs(O,F) and MgO by reducing the deposition temperature of NdOF over-layer needed for F-doping. The resultant NdFeAs(O,F) bicrystal films have a critical angle of 8.5$^\circ$, above which $J_{\rm c}$ starts to decrease exponentially. This critical angle is almost the same as those of other Fe-based superconductors.
\end{abstract}

\maketitle

\section{Introduction}
The discovery of iron-based superconductors (FBS) triggered a huge excitement for the community of fundamental and applied superconductivity research. Since then many physical quantities of FBS have been revealed, e.g. the coherence length is short, which results from the small Fermi velocity and low carrier density \cite{Putti}. The in-plane coherence length $\xi_{ab}(0)$ at zero kelvin obtained from upper critical field measurements is 2$\sim$4\,nm for FBS [e.g. the respective $\xi_{ab}(0)$ for Fe(Se,Te), BaFe$_2$(As$_{0.7}$As$_{0.3}$)$_2$, (Ba,K)Fe$_2$As$_2$, NdFeAsO$_{0.8}$F$_{0.2}$, and SmFeAs(O,F) were 1.5\,nm \cite{Klein}, 2.1\,nm \cite{Chaparro}, 1.6\,nm \cite{Kacmarcik}, 3.9\,nm \cite{Adamski}, and 1.4\,nm \cite{Welp}], which is similar to that for the cuprates. 
This short coherence length arouses concern that the superconductivity of FBS at grain boundaries may be depressed by crystalline disorder. However, the symmetry of the superconducting order parameter of FBS is reported to be an extended $s$-wave \cite{Mazin, Kuroki, Hanaguri}, which differs from the $d$-wave of the cuprates \cite{Tsuei}. Hence, the critical current density ($J_{\rm c}$) of FBS across the GB is expected to be not as severely reduced as cuprates. In fact, the critical angle ($\theta_{\rm c}$), above which $J_{\rm c}$ starts to decrease exponentially, is around 9$^\circ$ for Co-doped BaFe$_2$As$_2$ (Ba-122) \cite{Katase} and Fe(Se,Te) \cite{Si, Sarnelli-2}, much larger than YBa$_2$Cu$_3$O$_7$ ($\theta_{\rm c}$=3$^\circ$$\sim$5$^\circ$) \cite{Hilgenkamp}. Additionally, the grain boundaries for Co- and P-doped Ba-122 \cite{Katase, Sakagami}, and Fe(Se,Te) \cite{Si, Sarnelli-2} are of metallic nature. Both of these features are favourable for high-field conductor applications.

Among the various FBS $Ln$FeAs(O,F) ($Ln$: lanthanoide) has the highest superconducting transition temperature ($T_{\rm c}$), which yields a large margin between $T_{\rm c}$ and the temperature where a cryocooler can reach.
Hence, SmFeAs(O,F) wires \cite{Zhang} and NdFeAs(O,F) tapes \cite{Iida-1} have been demonstrated as proof-of-principle studies for conductor applications. However, a large $J_{\rm c}$ gap between those conductors and the counterpart of single crystals or epitaxial thin films was recognised \cite{Iida-2}. To understand the reason for such performance gap, GB characteristics of $Ln$FeAs(O,F) should be clarified.

Previously, we have investigated the grain boundary angle ($\theta_{\rm GB}$) dependence of inter-grain $J_{\rm c}$ ($J_{\rm c}^{\rm GB}$) for NdFeAs(O,F) grown on [001]-tilt symmetric MgO bicrystal substrates \cite{Omura}. As a result, $J_{\rm c}^{\rm GB}$ for $\theta_{\rm GB}$=6$^\circ$ was reduced by nearly 30\% compared to the intra-grain $J_{\rm c}$ ($J_{\rm c}^{\rm Grain}$). Microstructural investigation by transmission electron microscope (TEM) revealed that fluorine preferentially diffused through GB and thereby eroded NdFeAs(O,F) and the substrate. Hence, suppressing the excess diffusion of F is the key to exploring the intrinsic grain boundary properties. In the aforementioned study, F-doping has been conducted by a NdOF over-layer method \cite{Kawaguchi-2}, where parent NdFeAsO was deposited at 800$^\circ$C, followed by the deposition of NdOF at the same temperature. Here, a growth temperature of 800$^\circ$C was necessary for achieving high crystallinity of NdFeAsO \cite{Chihara}. On the other hand, the deposition temperature ($T_{\rm dep}$) of NdOF would not affect the crystalline quality of NdFeAs(O,F). Hence, it is possible to suppress the excess F diffusion by lowering $T_{\rm dep}$ of NdOF. However, the lack of F leads to low-$T_{\rm c}$ or even non-superconducting NdFeAs(O,F) films. 
Therefore, NdFeAs(O,F) bicrystal films should be fabricated by employing a $T_{\rm dep}$ of NdOF as low as possible without compromising the superconducting properties.

In this paper, we firstly determine the lowest $T_{\rm dep}$ for the NdOF over-layer that results into a NdFeAs(O,F) film with a sufficiently high $T_{\rm c}$ by mapping the diagram of $T_{\rm c}$ versus $T_{\rm dep}$. Then NdFeAs(O,F) bicrystal films are fabricated by employing the thus determined $T_{\rm dep}$ for NdOF, followed by structural characterisations as well as transport measurements.

\section{Experiment}
NdFeAs(O,F) thin films were grown on MgO(001) single crystalline substrates by molecular beam epitaxy (MBE) using solid sources of Fe, Fe$_2$O$_3$, As, NdF$_3$ and Ga. Here, Ga was used as a F-getter to adjust the amount of fluorine \cite{Kawaguchi-1}. A two-step process was employed to grow superconducting NdFeAs(O,F) thin films: parent NdFeAsO films of 40\,nm thickness were grown at 800$^\circ$C, followed by the deposition of 12\,nm thick NdOF in the temperature range $600^\circ{\rm C}\leq T_{\rm dep} \leq 800^\circ$C. For identifying the lowest deposition temperature of NdOF, $T_{\rm c}$ of the resultant NdFeAs(O,F) films was plotted as a function of $T_{\rm dep}$ (see fig.\,\ref{fig:figure2}). After determination of the lowest $T_{\rm dep}$ of NdOF, NdFeAs(O,F) thin films were fabricated on [001]-tilt symmetric MgO bicrystal substrates ($\theta_{\rm GB}$=6$^\circ$, 9$^\circ$, 12$^\circ$, 24$^\circ$) by the two-step process, whereas the respective thicknesses of NdFeAsO and NdOF were 160\,nm and 50\,nm to maintain the thickness ratio of NdOF to NdFeAsO as around 0.3.

Phase purity and $c$-axis texture were examined by x-ray diffraction (XRD) in Bragg-Brentano geometry using Cu-K$\alpha$ radiation. In-plane orientation of NdFeAs(O,F) was investigated by the $\phi$-scan of the 102 peak. After structural characterisation by XRD, the NdOF over-layer was removed by Ar-ion milling. Microstructural analyses on the NdFeAs(O,F) bicrystal film with $\theta_{\rm GB}$=6$^\circ$ were performed by TEM. From the TEM image of cross-sectional view away from the GB region, the thickness of NdFeAs(O,F) was found to be around 130\,nm after Ar-ion milling.

For transport measurements of both inter- and intra-grain $J_{\rm c}$, two micro-bridges were fabricated on the same film. The thin films were patterned by a photolithography method and then etched by Ar-ion milling to form micro-bridges.     
The bridges were 20-40\,\textmu${\rm m}$-wide and 1\,mm-long across the grain boundary and 0.25\,mm-long away from the grain boundary. Transport properties using the resultant bridges were measured by a four-probe method. The onset $T_{\rm c}$ was determined as the intersection between the fit to the normal state resistivity and the steepest slope of resistivity. An electric field criterion of $E$=1\,\textmu${\rm V/cm}$ was applied for evaluating intra-grain $J_{\rm c}$ ($J_{\rm c}^{\rm Grain}$) and inter-grain $J_{\rm c}$ ($J_{\rm c}^{\rm GB}$) for $\theta_{\rm GB}$=6$^\circ$ and 9$^\circ$. $J_{\rm c}^{\rm GB}$ for $\theta_{\rm GB}$=12$^\circ$ was defined as the intersection between $E=0$ and a linear fit to the non-ohmic linear differential (NOLD) region (i.e. the region where $E$ increased linearly with the current density $J$).

\section{Results and discussion} 
\begin{figure}[ht]
	\centering
		\includegraphics[width=14cm]{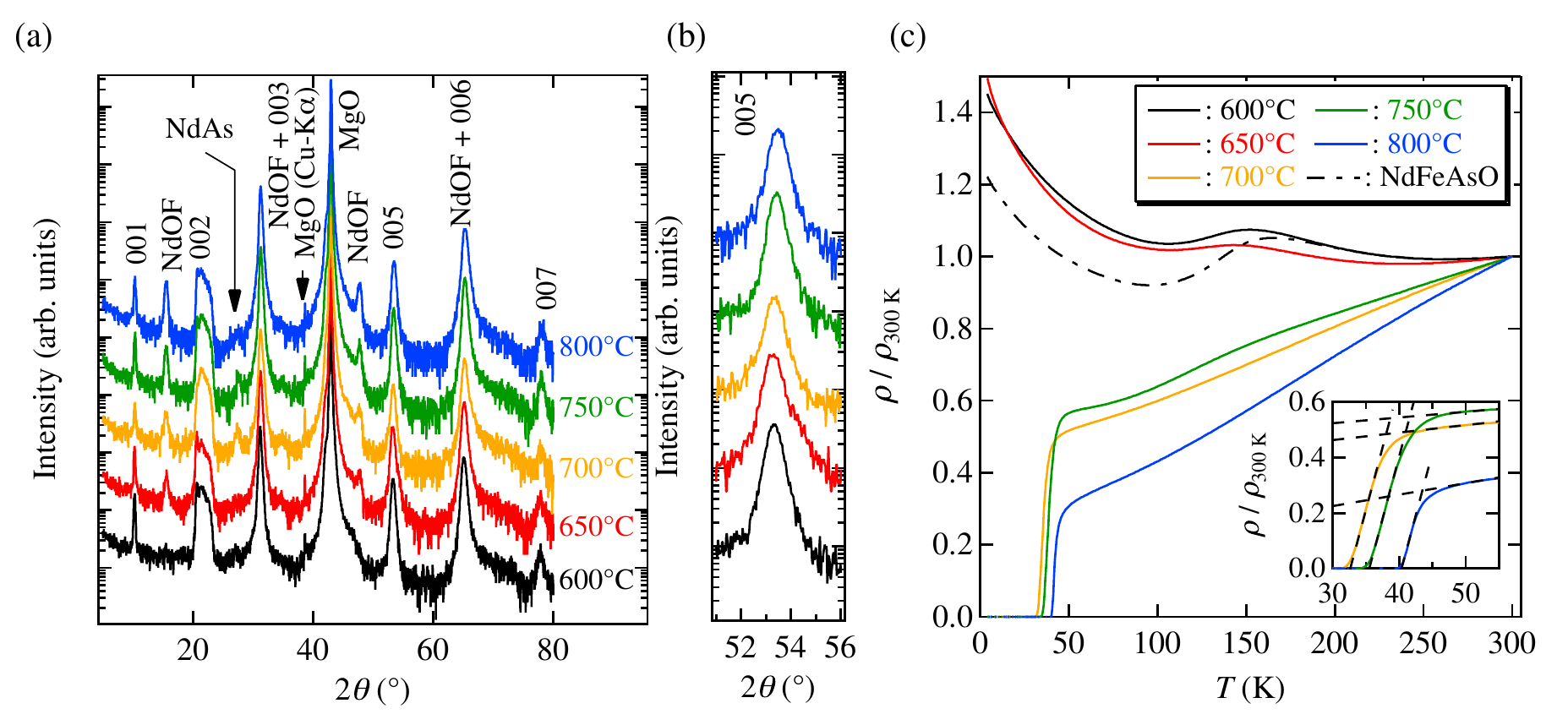}
		\caption{(a) The $\theta/2\theta$-scans (Cu-K$\alpha$) of the NdFeAs(O,F) thin films for various deposition temperatures of the NdOF over-layer. (b) Enlarged view of the 005 reflection. (c) The normalised resistivity curves of the NdFeAs(O,F) thin films for various deposition temperatures of the NdOF over-layer. For comparison, the data of the parent NdFeAsO film is also plotted. The data were normalised to the value of 300\,K. Inset shows the enlarged view of the superconducting transition. The dashed lines are fits to the normalised traces in the region of the transition and the normal states.} 
\label{fig:figure1}
\end{figure}

Figure\,\ref{fig:figure1}(a) shows the XRD patterns of the NdFeAs(O,F) thin films for various deposition temperatures of the NdOF over-layer. All  NdFeAs(O,F) films were grown $c$-axis oriented. The diffraction peak of NdAs, which is a by-product of chemical reaction between NdOF and NdFeAsO, appeared above $T_{\rm dep}$=700$^\circ$C. This is a natural consequence that higher deposition temperatures stimulate the chemical reaction. Another feature is a shift of the 00$l$ reflections to higher angles with increasing $T_{\rm dep}$, indicative of shortening the $c$-axis length [fig.\,\ref{fig:figure1}(b)]. It was reported that the $c$-axis lattice parameter of LaFeAsO$_{1-x}$F$_x$ \cite{Huang} and CeFeAsO$_{1-x}$F$_x$ \cite{Zhao} monotonously decreased with increasing F content. Hence, the observed change indicates that F diffuses into NdFeAsO and the amount of F in NdFeAs(O,F) increased with increasing the deposition temperature of the NdOF over-layer.

The temperature dependence of the normalised resistivity ($\rho/\rho_{\rm 300\,K}$) varies with $T_{\rm dep}$ [fig.\,\ref{fig:figure1}(c)]. Below $T_{\rm dep}$=650$^\circ$C the normalised resistivity curves of the films were almost similar to those of the parent NdFeAsO. The upturn of $\rho/\rho_{\rm 300\,K}$ at around 150\,K corresponds to the structural transition from a high temperature tetragonal phase to a low temperature orthorhombic one. An onset $T_{\rm c}$ of 38\,K was recorded for the film with $T_{\rm dep}$=700$^\circ$C. Further increase of the $T_{\rm dep}$ leads to an improvement of $T_{\rm c}$.

\begin{figure}[ht]
	\centering
		\includegraphics[width=8cm]{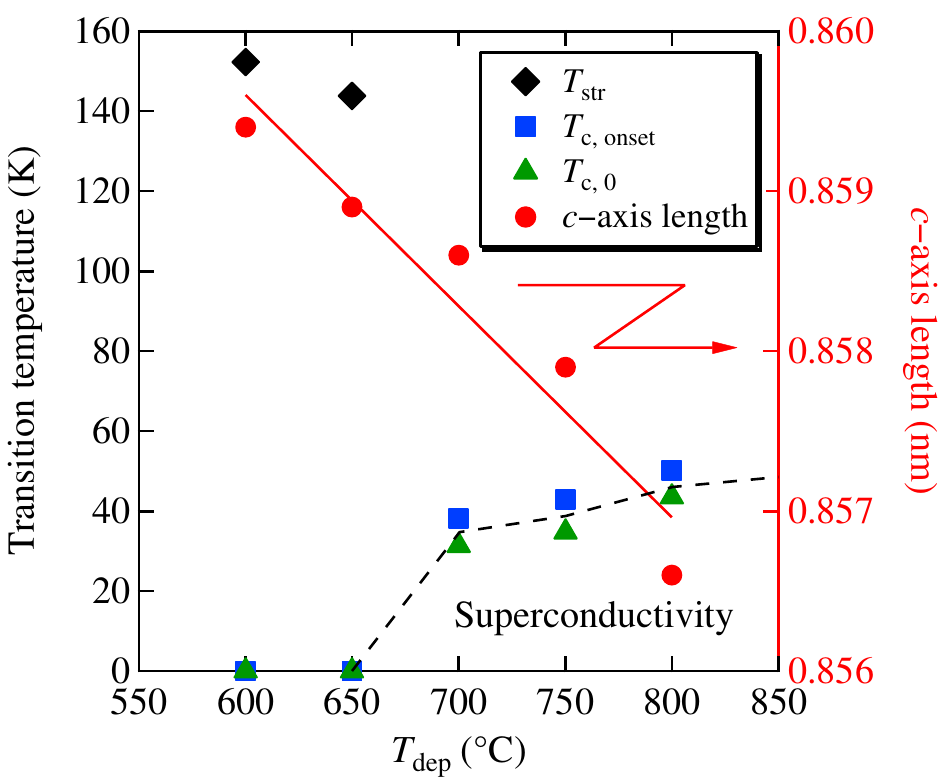}
		\caption{The transition temperatures and the $c$-axis lattice parameters for the NdFeAs(O,F) films as a function of the deposition temperature ($T_{\rm dep}$) of NdOF. The black diamond, blue square, green triangle, and red circle represent the structural transition temperature ($T_{\rm str}$), the onset superconducting transition temperature ($T_{\rm c, onset}$), the zero resistance temperature ($T_{\rm c, 0}$), and the $c$-axis length, respectively. Note that the films with $T_{\rm dep}$=600$^\circ$C and 650$^\circ$C did not show superconductivity. Red and dashed lines are guides for the eye.}
\label{fig:figure2}
\end{figure}

Figure\,\ref{fig:figure2} summarises the $c$-axis lattice parameter and $T_{\rm c}$ for the NdFeAs(O,F) films as a function of deposition temperature of NdOF. The $c$-axis lattice parameter decreased almost linearly with increasing $T_{\rm dep}$. Superconductivity can be induced by the deposition of NdOF at $T_{\rm dep}>650^\circ{\rm C}$. From these results, the lowest $T_{\rm dep}$ with keeping superconductivity is determined to be 700$^\circ$C.
In the following results, the NdOF over-layer was deposited at 700$^\circ$C.

\begin{figure}[t]
	\centering
		\includegraphics[width=10cm]{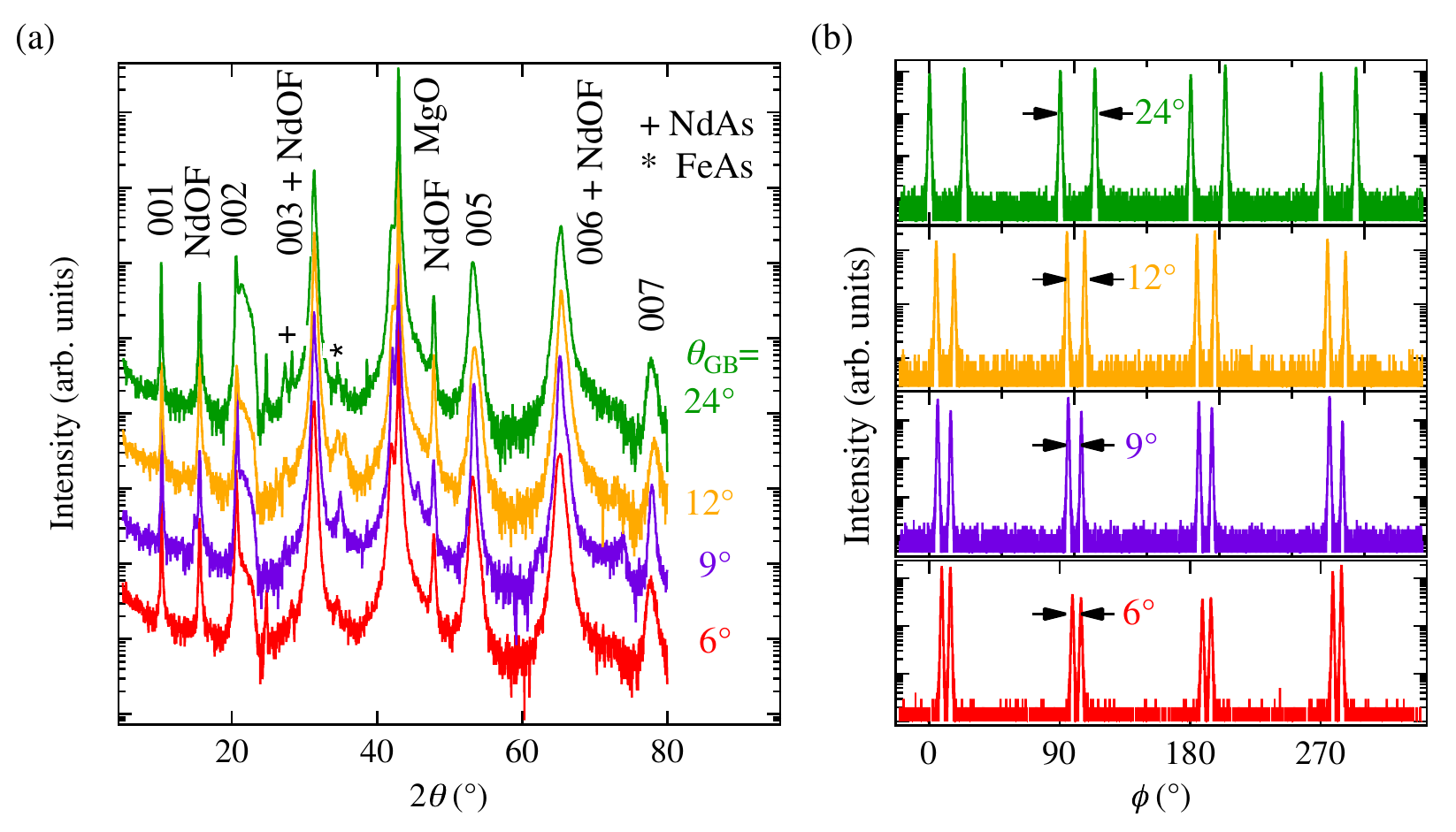}
		\caption{Summary of the structural characterisation for the NdFeAs(O,F) thin films fabricated on [001]-tilt symmetric MgO bicrystal substrates having various $\theta_{\rm GB}$. (a) The $\theta$/2$\theta$-scans and (b) the azimuthal $\phi$-scans of the off-axis 102 reflection of NdFeAs(O,F).}
\label{fig:figure3}
\end{figure}

Structural characterisation of the NdFeAs(O,F) films grown on [001]-tilt symmetric MgO bicrystal substrates is summarised in fig.\,\ref{fig:figure3}. The $\theta$/2$\theta$-scans showed that the NdFeAs(O,F) films were $c$-axis oriented. The azimuthal $\phi$-scan of the off-axis 102 reflection of NdFeAs(O,F)  showed clearly eight peaks from two adjacent NdFeAs(O,F) grains separated by $\theta_{\rm GB}$ [fig.\,\ref{fig:figure3}(b)]. Hence, the NdFeAs(O,F) films were grown epitaxially on [001]-tilt symmetric MgO bicrystal substrates.

\begin{figure}[ht]
	\centering
		\includegraphics[width=14cm]{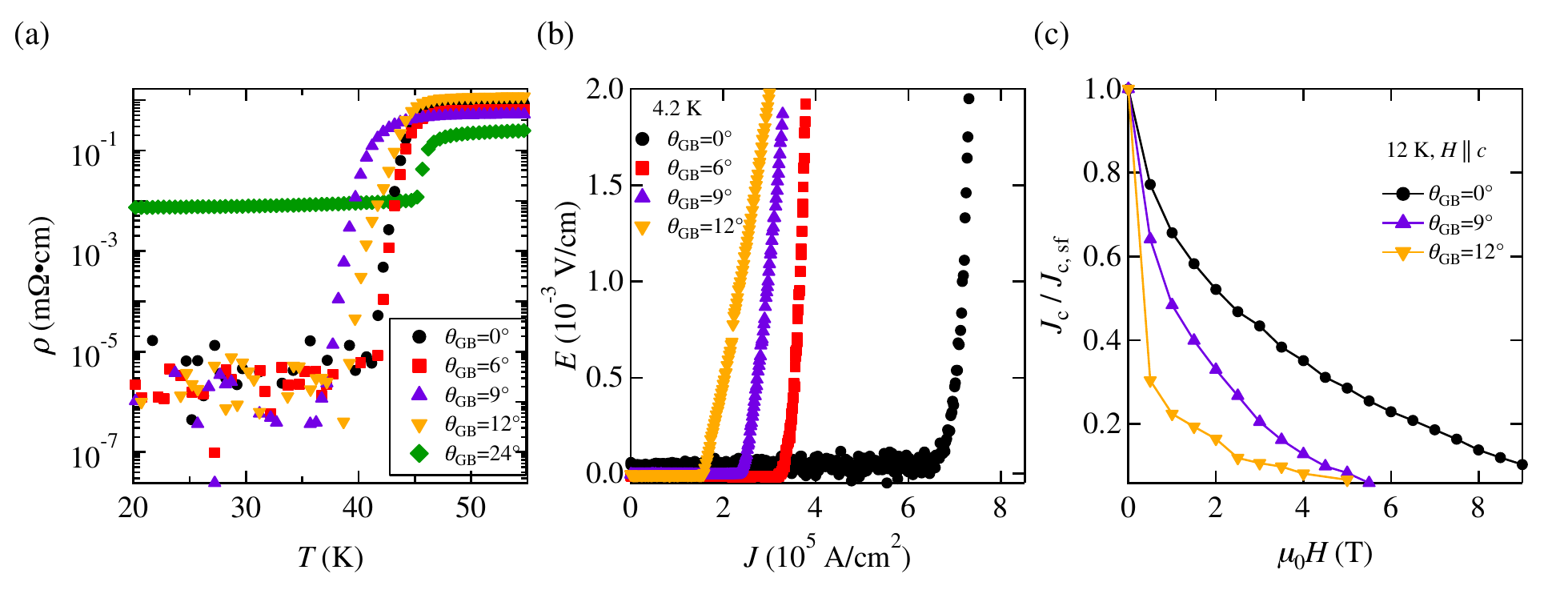}
		\caption{(a) Semi-logarithmic plot of the resistivity curves for the inter-grain bridges having various $\theta_{\rm GB}$ and the intra-grain bridge ($\theta_{\rm GB}$=0$^\circ$). The intra-grain bridge was fabricated from the NdFeAs(O,F) film with $\theta_{\rm GB}$=6$^\circ$. The measurements were conducted using 30\,\textmu${\rm m}$-wide bridges. (b) $E-J$ curves of the inter-grain bridges of 30\,\textmu${\rm m}$-wide for various $\theta_{\rm GB}$ and the intra-grain bridge ($\theta_{\rm GB}$=0$^\circ$) measured at 4.2\,K without magnetic field. The intra-grain bridge was fabricated from the NdFeAs(O,F) film with $\theta_{\rm GB}$=12$^\circ$. (c) Comparative field dependence of $J_{\rm c}$ normalised to the self-field $J_{\rm c}$ ($J_{\rm c, sf}$) measured at 12\,K for $H$$\parallel$$c$. The bridges are the same as those presented in (b).} 
\label{fig:figure4}
\end{figure}

In fig.\,\ref{fig:figure4}(a) the resistivity curves for the inter-grain bridges (bridge width $w$=30\,\textmu${\rm m}$) with various misorientation angles are shown. For comparison the data of the intra-grain bridge ($\theta_{\rm GB}$=0$^\circ$) fabricated from the NdFeAs(O,F) film with $\theta_{\rm GB}$=6$^\circ$ are also plotted. As can be seen, the superconducting transition temperature did not systematically change with the grain boundary angle, which differs from our previous study (i.e. $T_{\rm c}$ decreased with increasing $\theta_{\rm GB}$) \cite{Omura}. 
The inter-grain bridges with $\theta_{\rm GB}$=6$^\circ$ and 12$^\circ$ had an onset $T_{\rm c}$ ($T_{\rm c, onset}$) of 45\,K, which is almost the same as for the intra-grain bridges. For $\theta_{\rm GB}\geq9^\circ$ a wide transition width of 5\,K was observed. As shown in Supplementary fig.\,S1, almost no difference in  $T_{\rm c}$ as well as the superconducting transition width were found with changing the bridge width.

A finite resistivity was observed for the inter-grain bridge with $\theta_{\rm GB}$=24$^\circ$ below $T_{\rm c}$, whereas the other bridges showed a resistivity well below the instrumental limitation. Hence, the grain boundary region for $\theta_{\rm GB}$=24$^\circ$ was destroyed by F even at $T_{\rm dep}$=700$^\circ{\rm C}$, and accordingly, $J_{\rm c}$ across the GB is zero.

Figure\,\ref{fig:figure4}(b) compares the $E-J$ curves for the inter-grain bridges with various $\theta_{\rm GB}$ measured at 4.2\,K without a magnetic field. The curve for the intra-grain bridge ($\theta_{\rm GB}$=0$^\circ$) is also included, which exhibited a power-law behaviour described by $E\sim J^n$, representing flux creep effects. The same behaviour was also seen for the inter-grain bridges with $\theta_{\rm GB}$=6$^\circ$ and 9$^\circ$, which indicates that the two adjacent grains were strongly coupled. On the other hand, the inter-grain curve for $\theta_{\rm GB}$=12$^\circ$ showed a non-Ohmic linear differential (NOLD) behaviour (i.e. $E$ is linearly changing with $J$) originating from viscous flux flow along the grain boundaries \cite{Diaz}, indicative of $J_{\rm c}$ limitation by grain boundaries \cite{Verebelyi}. The resistance area product for $\theta_{\rm GB}$=12$^\circ$ was estimated to be 8.5$\times$10$^{-10}$\,$\Omega$cm$^2$, which is of the same order of magnitude for Co-doped Ba-122 with $\theta_{\rm GB}$=45$^\circ$ (5$\times$10$^{-10}$\,$\Omega$cm$^2$)\cite{Katase}, P-doped Ba-122 with $\theta_{\rm GB}$=24$^\circ$ (1.3$\times$10$^{-10}$\,$\Omega$cm$^2$)\cite{Sakagami}, and Fe(Se,Te) with $\theta_{\rm GB}$=24$^\circ$ (7$\times$10$^{-10}$\,$\Omega$cm$^2$)\cite{Si}. Hence, the grain boundary of NdFeAs(O,F) is of metallic nature, which is similar to other FBS.

The field dependence of $J_{\rm c}$ normalised to the self-field $J_{\rm c}$ ($J_{\rm c, sf}$) plotted in fig.\,\ref{fig:figure4}(c) also shows a typical behaviour of GB transport observed for FBS. For the bridge with $\theta_{\rm GB}$=12$^\circ$, inter-grain $J_{\rm c}$ was decreased by 70\% even at a small applied magnetic field of 0.5\,T, whereas the corresponding reduction for $\theta_{\rm GB}$=0$^\circ$ (i.e. intra-grain) and $\theta_{\rm GB}$=9$^\circ$  were 23\% and 36\%, respectively. Such a sharp drop of inter-grain $J_{\rm c}$ for large $\theta_{\rm GB}$ by small magnetic fields is due to the weak coupling nature between adjacent grains.

\begin{figure}[t]
	\centering
		\includegraphics[width=15cm]{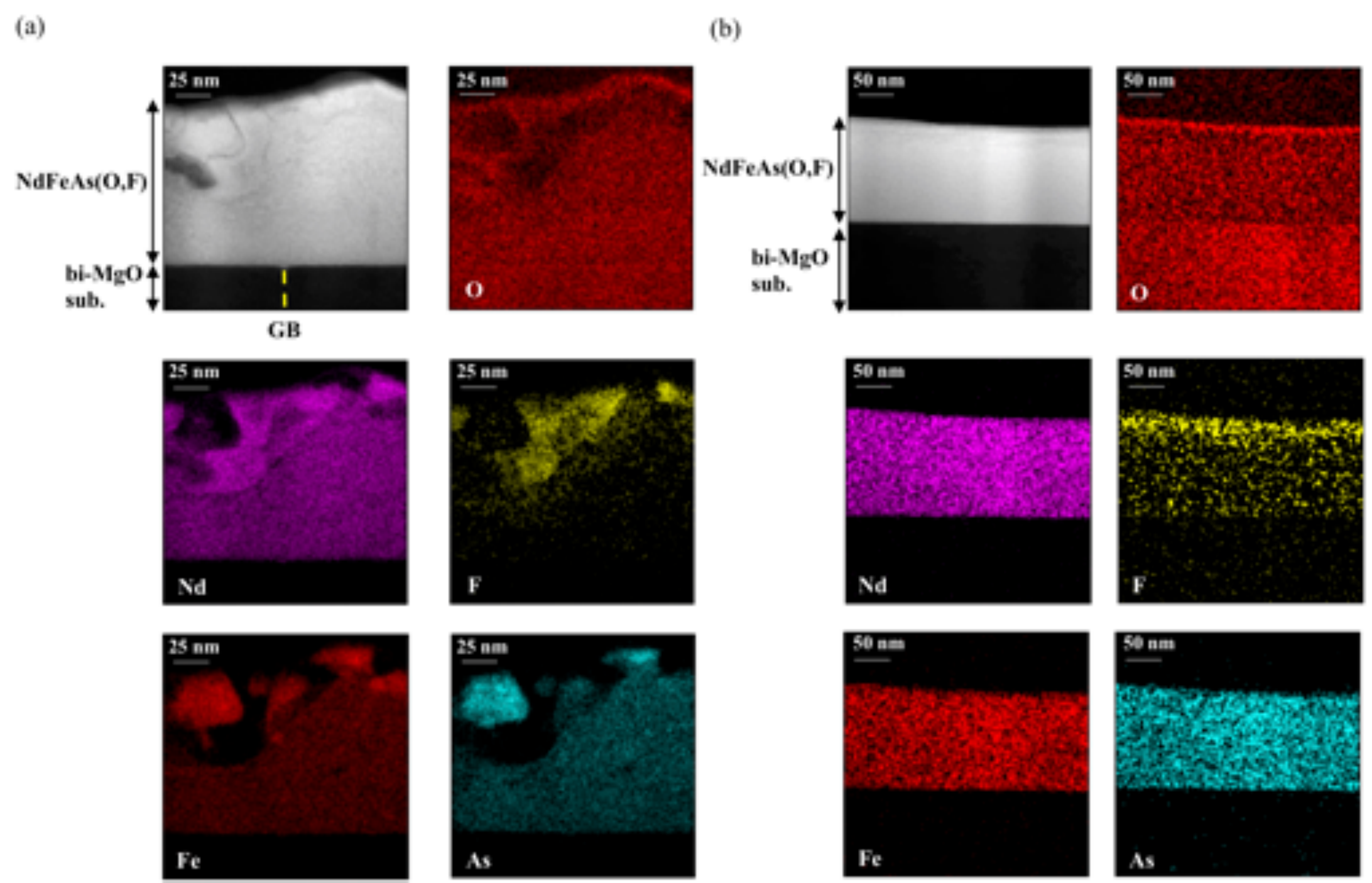}
		\caption{(a) TEM micrograph of a cross section of the GB region having $\theta_{\rm GB}$=6$^\circ$ and the elemental mappings at the same area. The location of the GB is indicated by a dashed yellow line in the micrograph. (b) The microstructures away from the GB region. The thickness of NdFeAs(O,F) was around 130\,nm.}
\label{fig:figure5}
\end{figure}

To further characterise the coupling between the grains with $\theta_{\rm GB}$=6$^\circ$, microstructural observation by TEM was conducted [fig.\,\ref{fig:figure5}(a)]. For comparison, the microstructures away from the GB region are also shown in fig.\,\ref{fig:figure5}(b).
The elemental mappings revealed a uniform distribution of all elements except for O and F [fig.\,\ref{fig:figure5}(b)]. The observed high concentration of O and F at the surface region may be caused by heating due to Ar-ion milling to remove the NdOF over-layer. Uniform distribution of Nd, Fe, As, O and F within NdFeAs(O,F) was also observed in a different bridge (see Supplementary fig.\,S2). In stark contrast, the elemental distribution of the top region of NdFeAs(O,F) around the GB was strongly inhomogeneous due to the damage to NdFeAs(O,F) by F even though the deposition temperature of the NdOF over-layer was lowered [fig.\,\ref{fig:figure5}(a)]. This is very similar to what was observed in our previous investigation\cite{Omura}. However, the bottom region (about half of the thickness from the substrate) was quite uniform and free from damage by F. Hence, the grains are still coupled strongly in this region.
     
\begin{figure}[b]
	\centering
		\includegraphics[width=15cm]{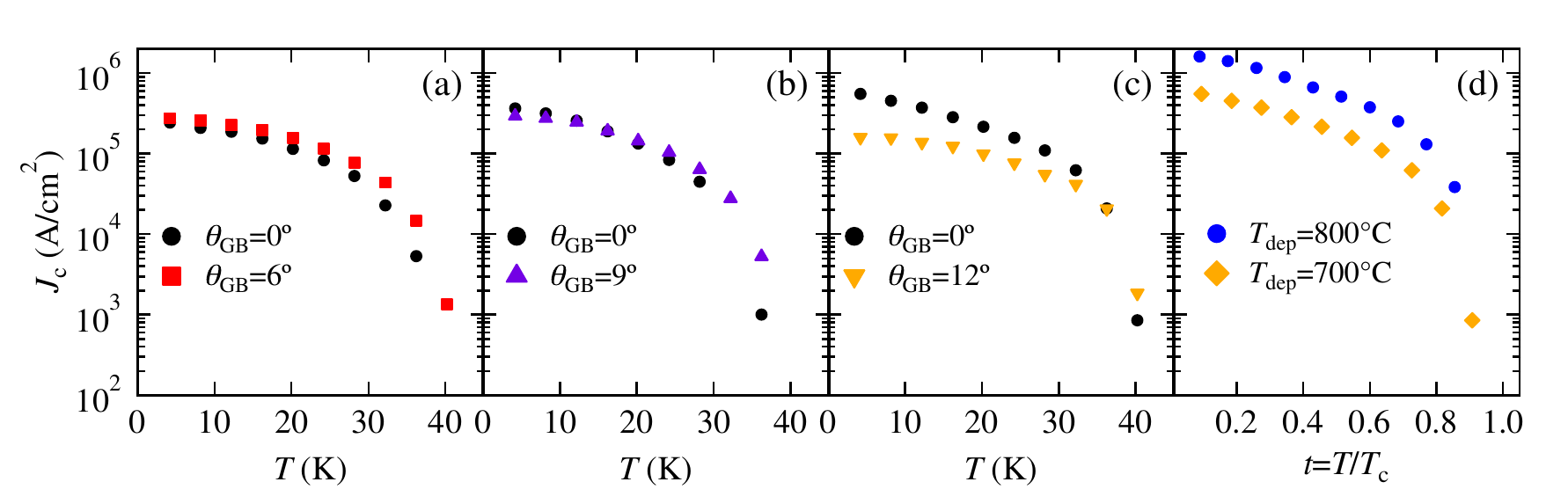}
		\caption{(a)--(c) Temperature dependence of the self-field $J_{\rm c}$ for the intra- and inter-grain bridges (bridge width $w$=30\,\textmu${\rm m}$) having various $\theta_{\rm GB}$. (d) Self-field $J_{\rm c}^{\rm Grain}$ as a function of reduced temperature for NdFeAs(O,F) films grown by employing $T_{\rm dep}$=700$^\circ {\rm C}$ and 800$^\circ {\rm C}$.}
\label{fig:figure6}
\end{figure}

The temperature dependence of self-field $J_{\rm c}$ for various misorientation angles are summarised in figs.\,\ref{fig:figure6}(a)--(c). The intra-grain $J_{\rm c}$ ($J_{\rm c}^{\rm Grain}$) of each film is also shown for comparison. Here, the measurements were conducted using 30\,\textmu${\rm m}$-wide bridges. The results obtained from various bridges are summarised in Supplementary fig.\,S3. The self-field $J_{\rm c}$ of the intra-grain at low temperatures is in the order of $10^5$\,A/cm$^2$ for all three samples, which is one order of magnitude lower than our NdFeAs(O,F) films grown by employing $T_{\rm dep}$=800$^\circ {\rm C}$ \cite{Chiara}. By comparing the $J_{\rm c}^{\rm Grain}$ at the same reduced temperatures ($t=T/T_{\rm c}$), $J_{\rm c}^{\rm Grain}$ of an ordinal NdFeAs(O,F) film prepared with $T_{\rm dep}$=700$^\circ {\rm C}$ is smaller than that of a film with $T_{\rm dep}$=800$^\circ {\rm C}$ [fig.\,\ref{fig:figure6}(d)]. This is due to low carrier concentration. Hall effect measurements on our NdFeAs(O,F) films grown on single crystalline MgO revealed that the carrier concentration  decreased from 1.98$\times$10$^{21}$\,/cm$^3$ for $T_{\rm dep}$=800$^\circ {\rm C}$ to 1.34$\times$10$^{21}$\,/cm$^3$ for $T_{\rm dep}$=750$^\circ {\rm C}$ at 50\,K \cite{Matsumoto}, although a precise evaluation of the carrier concentration is not easy due to the multi-band nature of the current superconductor. The carrier concentration of NdFeAs(O,F) for $T_{\rm dep}$=700$^\circ {\rm C}$ is expected to be even smaller than that for $T_{\rm dep}$=750$^\circ {\rm C}$, which probably explains the smaller $J_{\rm c}$.

The distinct feature of the above results is that the respective inter-grain critical current densities ($J_{\rm c}^{\rm GB}$) for $\theta_{\rm GB}$=6$^\circ$ and $9^\circ$ are almost comparable to their intra-grain $J_{\rm c}$ ($J_{\rm c}^{\rm Grain}$) at all temperatures [figs.\,\ref{fig:figure6}(a), (b)], indicative of the absence of weak-links for both bicrystal films. For $\theta_{\rm GB}$=12$^\circ$ $J_{\rm c}^{\rm GB}$ below 32\,K was certainly lower than $J_{\rm c}^{\rm Grain}$. In Cu-Ag alloys it has been observed that the GBs are completely wetted by Ag when the misorientation angle is large \cite{Straumal}. It might be possible that F-rich phase had covered GBs and caused the transition from strong-link to weak-link behaviour. Nevertheless, $J_{\rm c}^{\rm GB}$ at 4.2\,K decreased by only 28\% compared to $J_{\rm c}^{\rm Grain}$ in stark contrast to our previous study \cite{Omura}. Note that the effective cross-sectional area in the vicinity of the GB with $\theta_{\rm GB}$=6$^\circ$ was smaller than the intra-grain bridge as a consequence of the damage caused by F diffusion [see fig.\,\ref{fig:figure5}(a)]. This is probably the reason why the plotted $J_{\rm c}^{\rm GB}$ values are  higher than $J_{\rm c}^{\rm Grain}$ in fig.\,\ref{fig:figure6}(a), since $J_{\rm c}$ was calculated by assuming that both the inter- and intra-grain bridges have the identical thickness.

\begin{figure}[t]
	\centering
		\includegraphics[width=7cm]{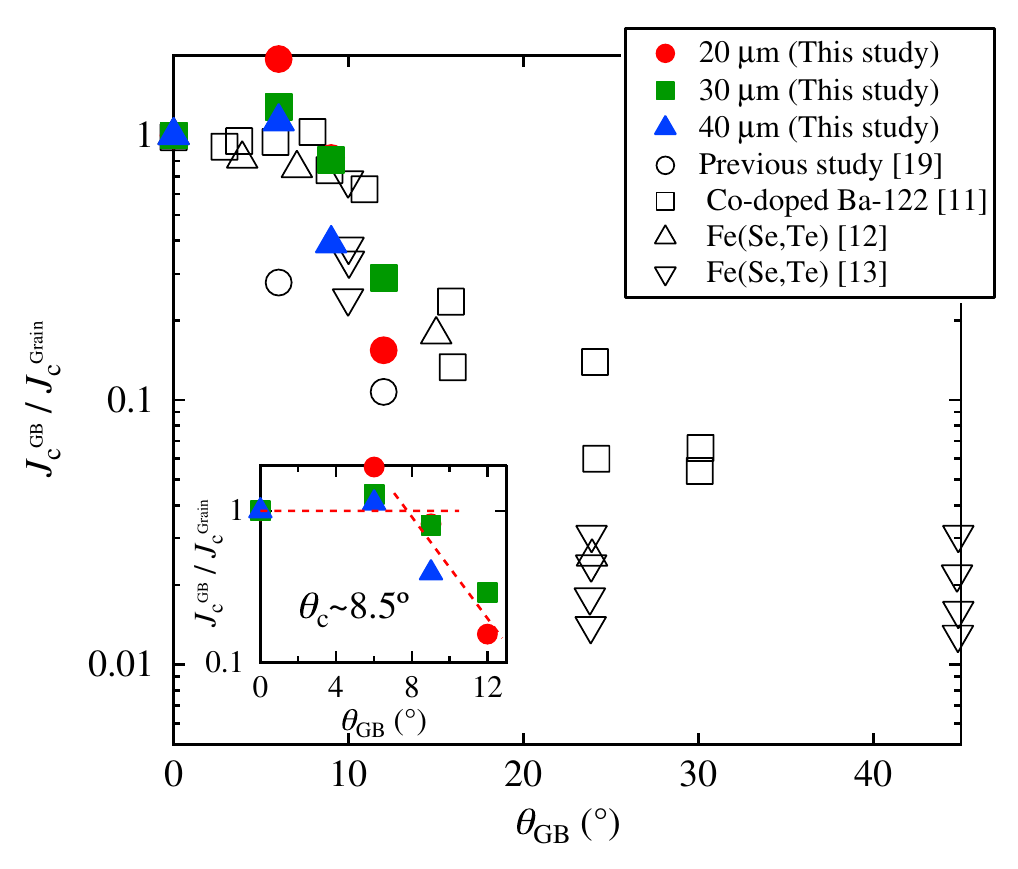}
		\caption{The ratio of inter- to intra-grain $J_{\rm c}$ ($J_{\rm c}^{\rm GB}/J_{\rm c}^{\rm Grain}$) at 4.2\,K for NdFeAs(O,F) bicrystal films as a function of $\theta_{\rm GB}$. The data of Co-doped Ba-122 at 4\,K \cite{Katase}, Fe(Se,Te) at 4.2\,K \cite{Si, Sarnelli-2}, and our previous study on NdFeAs(O,F) at 4.2\,K \cite{Omura} are also plotted for comparison. Inset shows an enlarged view in the vicinity of the critical angle ($\theta_{\rm c}$).}
\label{fig:figure7}
\end{figure}

Figure\,\ref{fig:figure7} shows the $\theta_{\rm GB}$ dependence of the ratio of inter- to intra-grain $J_{\rm c}$ ($J_{\rm c}^{\rm GB}/J_{\rm c}^{\rm Grain}$) for the NdFeAs(O,F) bicrystal films at 4.2\,K. For comparison the data of 
Co-doped Ba-122 at 4\,K \cite{Katase}, Fe(Se,Te) at 4.2\,K \cite{Si, Sarnelli-2}, and our previous work on NdFeAs(O,F) at 4.2\,K \cite{Omura} are also plotted. The present work on NdFeAs(O,F) bicrystal films showed that $J_{\rm c}^{\rm GB}/J_{\rm c}^{\rm Grain}$ was unity up to 8.5$^\circ$ (i.e. $\theta_{\rm c}=8.5^\circ$), followed by a decrease with $\theta_{\rm GB}$. On the other hand, $\theta_{\rm c}$ was less than 6$^\circ$ in our previous investigation. These results indicate that the grain boundary properties were improved by lowering $T_{\rm dep}$ of NdOF. It is interesting that all FBS have a similar critical angle of around 9$^\circ$, suggesting that the symmetry of the superconducting order parameter for NdFeAs(O,F) may be the same as Co-doped Ba-122 and Fe(Se,Te).

Although $\theta_{\rm c}$=8.5$^\circ$ was determined for NdFeAs(O,F), it is still ambiguous how the inter-grain $J_{\rm c}$ decreases with $\theta_{\rm GB}$ beyond 12$^\circ$.
Additionally, it is not ruled out completely that the damage to the GB region by F had influenced the obtained results. To answer these issues and explore further the intrinsic GB characteristics, studying F-free oxypnictides, e.g. $Ln$(Fe$_{1-x}$Co$_x$)AsO \cite{Wang, Kim}, would be interesting. Albeit the highest $T_{\rm c}$ is around 25\,K for Nd(Fe$_{1-x}$Co$_x$)AsO \cite{Kim}, there would be no damage to the GB region by F. The misorientation angle dependence of $J_{\rm c}^{\rm GB}$ for Nd(Fe$_{1-x}$Co$_x$)AsO bicrystal films would be the direction of our future studies.

\section{Summary} 
NdFeAs(O,F) epitaxial thin films were fabricated by MBE using a two-step process, where the deposition temperature ($T_{\rm dep}$) of NdOF was varied in the temperature range $600^\circ{\rm C}\leq T_{\rm dep} \leq 800^\circ$C. The $c$-axis lattice parameter and $T_{\rm c}$ of NdFeAs(O,F) changed systematically with $T_{\rm dep}$ of NdOF, and the F content in NdFeAs(O,F) increased with increasing $T_{\rm dep}$. The lowest $T_{\rm dep}$ of NdOF without compromising $T_{\rm c}$ was determined as 700$^\circ$C. By using $T_{\rm dep}$=700$^\circ{\rm C}$ epitaxial NdFeAs(O,F) thin films were grown on [001]-tilt symmetric MgO bicrystal substrates with $\theta_{\rm GB}$=6$^\circ$, 9$^\circ$, 12$^\circ$ and 24$^\circ$. Even for this reduced $T_{\rm dep}$, however, the inter-grain bridge with $\theta_{\rm GB}$=24$^\circ$ showed a finite resistance and hence $J_{\rm c}$ was zero due to the erosion of NdFeAs(O,F) by F. 
Additionally, $J_{\rm c}$ of the inter-grain bridges with $\theta_{\rm GB} \leq \theta_{\rm c}$ and the intra-grain bridge were around 10$^5$\,A/cm$^2$ at low temperatures, which is almost one order of magnitude lower than NdFeAs(O,F) with $T_{\rm dep}$=800$^\circ{\rm C}$ due probably to low carrier concentration. Nevertheless, the GB properties of the inter-grain bridges with lower $\theta_{\rm GB}$ were improved, resulting in a critical angle of $\theta_{\rm c}$=8.5$^\circ$. This critical angle is identical to Co-doped Ba-122 and Fe(Se,Te).
  
\ack
This work was supported by the JSPS Grant-in-Aid for Scientific Research (B) Grant Number 16H04646 as well as JST CREST Grant Number JPMJCR18J4.

\clearpage
\setcounter{figure}{0}
\renewcommand{\thefigure}{S\arabic{figure}}

\section*{Supplementary information}

\begin{figure}[h]
	\centering
		\includegraphics[width=16cm]{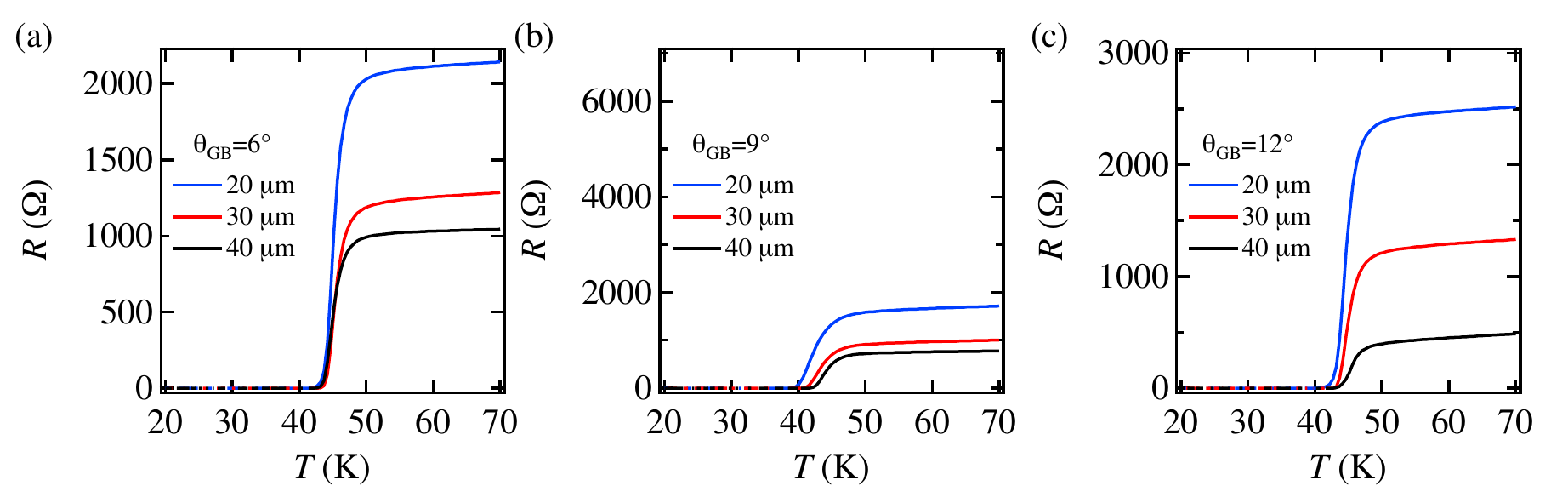}
		\caption{(Color online) Temperature dependence of the resistance curves for the inter-grain bridges, (a) $\theta_{\rm GB}$=6$^\circ$, (b) $\theta_{\rm GB}$=9$^\circ$ and (c) $\theta_{\rm GB}$=12$^\circ$, having various bridge width.} 
\label{fig:figureS1}
\end{figure}

\begin{figure}[h]
	\centering
		\includegraphics[width=16cm]{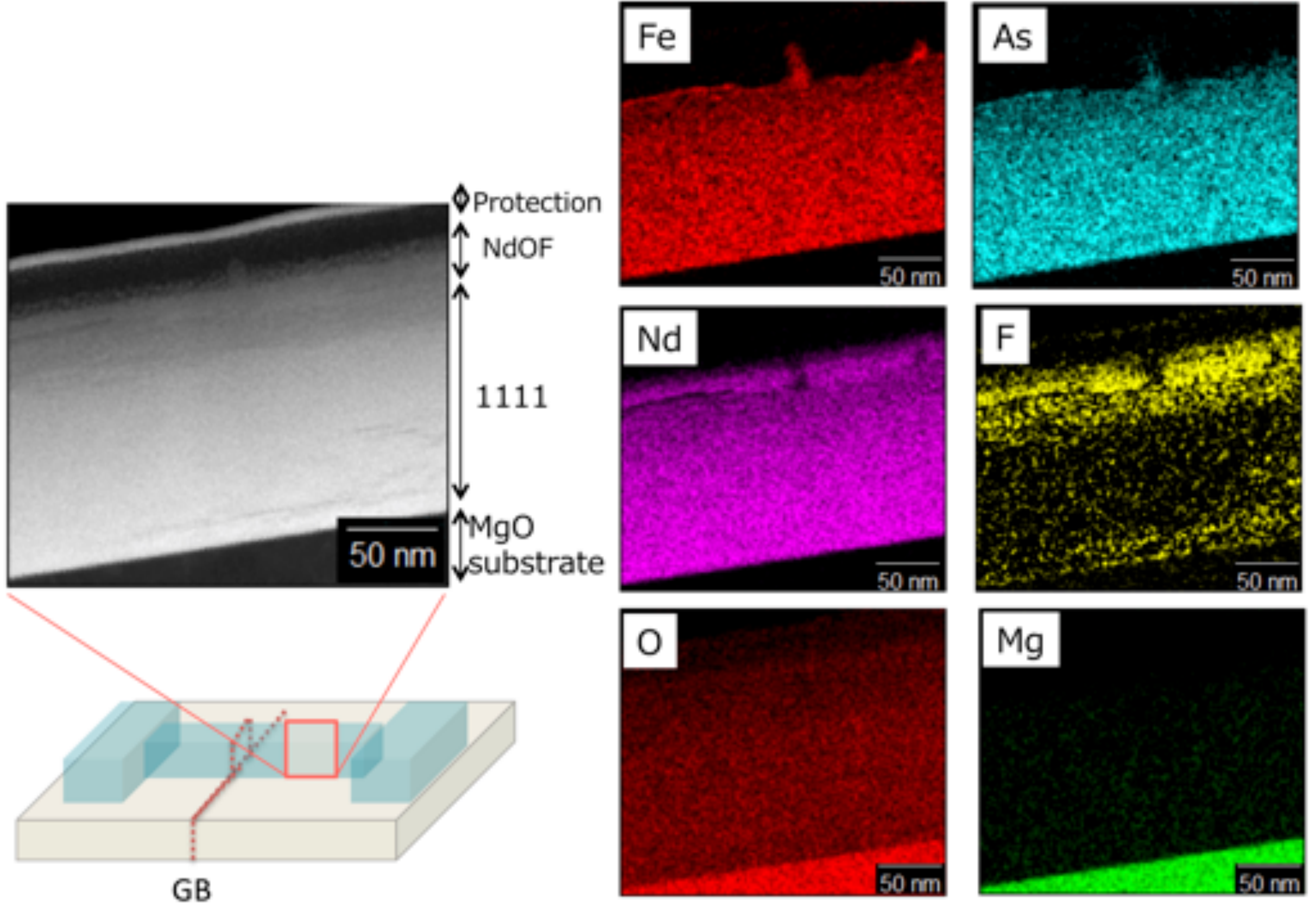}
		\caption{(Color online) Cross-sectional TEM micrograph and elemental mappings acquired at a position away from the GB region. The micro-bridge used for this observation is the one with ($\theta_{\rm GB}$=24$^\circ$) reported in our previous study \cite{Omura}. Note that the NdOF over-layer was not removed.} 
\label{fig:figureS2}
\end{figure}

\begin{figure}[h]
	\centering
		\includegraphics[width=16cm]{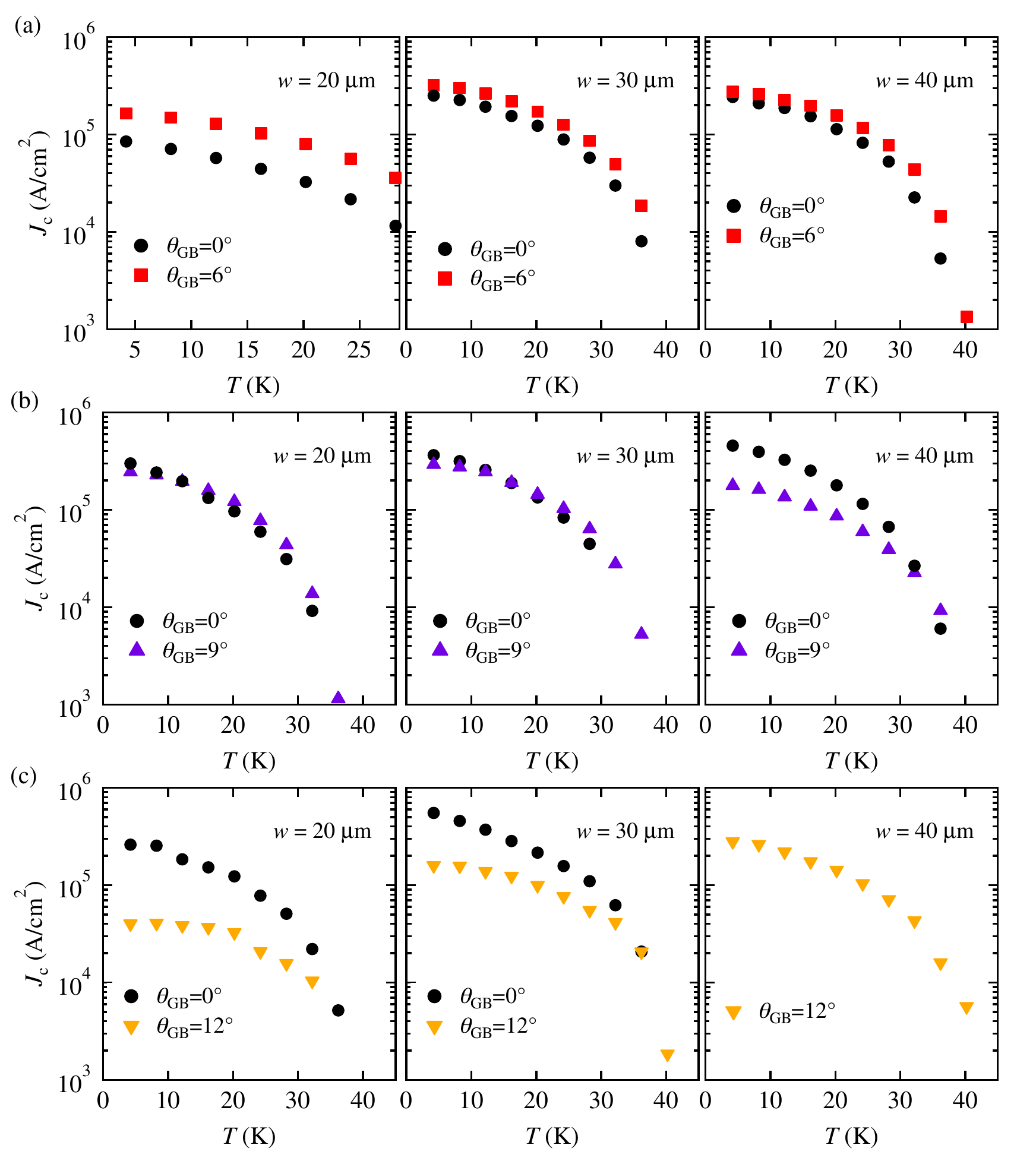}
		\caption{(Color online) Temperature dependence of the self-field Jc for the intra- ($\theta_{\rm GB}$=0$^\circ$) and inter-grain bridges, (a) $\theta_{\rm GB}$=6$^\circ$, (b) $\theta_{\rm GB}$=9$^\circ$ and (c) $\theta_{\rm GB}$=12$^\circ$, having various bridge width, $w$. For $\theta_{\rm GB}$=12$^\circ$ the 40\,\textmu${\rm m}$-wide bridge for intra-grain measurements was damaged and, therefore, the data was not available.} 
\label{fig:figureS3}
\end{figure}

\end{document}